\documentclass[usenatbib]{basi}
\usepackage[T1]{fontenc}
\usepackage[british]{babel}
\usepackage[varg]{txfonts}
\usepackage{rotating}
\usepackage{dcolumn}
\begin{document}
\title[MHD Oscillations of Stellar Coronae]{MHD Seismology as a Tool to Diagnose the Coronae of X-ray Active Sun-like Flaring Stars}
\author[A.~K.~Srivastava and S. Lalitha]%
       {A.~K.~Srivastava$^1$\thanks{email: \texttt{asrivastava.app@iitbhu.ac.in} (Present Address: Department of Physics, Indian Institute of Technology (BHU), Varanasi-221005, India.)},
       Sairam~Lalitha$^{2}$\\
       $^1$Aryabhatta Research Institute of Observational Sciences (ARIES), Manora Peak, Nainital-263 129, India.\\
       $^2$Hamburger Sternwarte, University of Hamburg, Gojenbergsweg 112, 21029, Hamburg, Germany.}

\pubyear{2013}
\volume{00}
\pagerange{\pageref{firstpage}--\pageref{lastpage}}

\date{Received --- ; accepted ---}

\maketitle
\label{firstpage}

\begin{abstract}
It is now well accepted that the detection of impulsively generated multiple MHD
modes are potentially used in diagnosing the local plasma conditions of the solar corona.
Analogously, such analyses can also be significantly used in diagnosing the coronae of X-ray active
Sun-like stars. In the present paper, we briefly review the detection of MHD modes in coronae of
some X-ray active Sun-like stars, e.g. Proxima Centauri, $\xi$-Boo etc using XMM-Newton observations, and discuss the implications in deriving physical information about their localized magnetic atmosphere.  
We conclude that
the refinement in the MHD seismology of solar corona is also providing the best analogy to
develop the stellar seismology of magnetically active and flaring Sun-like stars to deduce 
the local physical conditions of their coronae.
\end{abstract}


\section{Introduction}\label{s:intro}

The mega-Kelvin plasma coupled with the magnetic field generates variety of 
magnetohydrodynamic wave modes in the solar corona that are one of the 
important candidates in its heating and plasma dynamics (cf., Nakariakov \& 
Verwichte, 2005). The magnetic field activity generated in the sub-photospheric 
layers and fanning out in the outer atmosphere of the Sun, leads other variety 
of plasma processes (e.g., jets, surges, spicules, hot plasma 
flows etc) as well as transient 
phenomena (e.g., solar flares, eruptive filaments, CMEs etc) at diverse spatio-temporal 
scales, which are crucial to transport mass and energy (e.g., 
De Pontieu et al., 2004; Gopalswamy et al., 2012; De Pontieu et al,. 2011; Srivastava \& Murawski, 2011; 
Shibata \& Magara, 2011; Joshi et al. 2012a,b; Kayshap et al., 2013a,b, and references 
cited therein). Along with these transient plasma processes, the solar coronal magnetic 
fields play important role in yielding high energy particles, emissions over the 
whole electromagnetic spectrum from Gamma-rays, X-rays to the Radio waves, and infer 
about variety of physical processes of coronal magneto-plasma (e.g, Vilmer, 2012, and references cited therein).
These transients also trigger the tube-waves (e.g., kink, sausage, torsional, and slow modes),
which are very important candidates to derive the local plasma conditions (e.g., magnetic field)
of the solar corona (e.g., Nakariakov \& Ofman, 2001; Nakariakov \& Verwichte, 2005; 
Srivastava et al., 2008; Andries et al. 2009; Srivastava \& Dwivedi,
2010; Kumar et al., 2011; Luna-Cardozo et al. 2012, and references cited therein).

The Sun-like stars, especially the cool sub-dwarfs, having the 
sufficiently thick convection zone in their sub-surface layers, can transport 
the magnetic fields in their coronae, and exhibit the transients with 
high energy emissions (e.g., X-rays), coronal heating,  as well as some signature of the 
MHD waves and oscillations (e.g., Mitra-Kraev et al., 2005; Pandey \& Srivastava, 2009).
The solar MHD siesmology is now in very advanced state with the detection of various 
tubular MHD modes in the variety of magnetized structures at diverse spatio-temporal
scales, and being very useful on deducing the local plasma conditions (e.g., 
O'Shea et al., 2007; Srivastava et al., 2008; Verwichte et al., 2010; Srivastava et al., 2010; Kumar et al.,2011; Kim et al., 2012; Srivastava et al., 2013,
and references cited therein). However, 
the detection of transients (e.g., flares) induced MHD waves and oscillations is very few in the case of the stellar coronae of Sun-like stars (e.g., Mitra-Kraev et al., 2005;
Pandey \& Srivastava, 2009, Anfinogentov et al. 2013, and references cited there). Moreover, the determination of local 
plasma structuring and magnetic field conditions are mostly unknown in the case of stellar coronae , and the solar analogy 
of the MHD seismology can be useful in deducing such conditions to know more about the physical properties of the
stellar coronae (e.g., Nakariakov \& Ofman, 2001). In the present paper, we briefly
review the detection of MHD waves and oscillations in the X-ray active coronae 
of Sun-like flaring stars, as well as discuss their significance in the framework of
MHD seismology.


\begin{figure}
\centerline{\includegraphics[width=10cm]{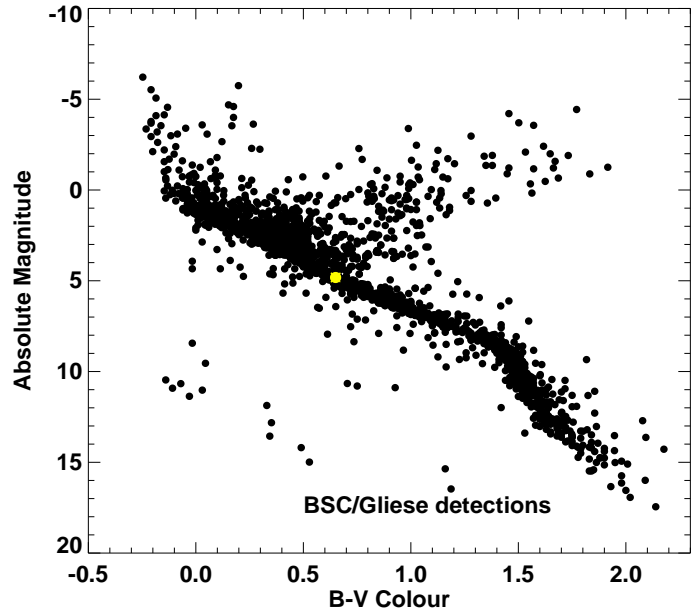}}
\caption{H-R diagram showing the Sun-like stars having X-ray activities and lying around the Sun (yellow circle). The cool 
stars around the Sun possessing the Sun-like magnetic coronae may have typical transient phenomena (e.g., flares)
as well as MHD processes
(Credit : Data from Huensch et al. 1997, 1998; Ph.D Thesis of S. Lalitha ).
\label{f:one}}
\end{figure}
\section{X-Ray Active Sun-like Flaring Stars}\label{X-ray  Stars}

The energy in stellar flares is originated from the interaction/reconnection
of magnetic fields that are permeating the coronae of such flaring stars and originally formed 
in their interior. The magnetic fields fan-out
through their convection zone into photosphere and finally into their coronae
to form the stellar magnetic loops.
The flares occur due to the magnetic field line reconnection in the coronae leading to the
heating and particle acceleration in the reconnection region as per the analogy 
of the occurrence of solar flares (e.g., see the reviews by Benz, 2008;
Shibata \& Magara 2011, and references cited therein). Particles
are accelerated downwards into the chromosphere to 
heat the denser ambient plasma, which further expands and evaporates into the
corona during the solar and stellar flaring processes. The stellar loops are filled with denser and hot plasma that is observed
in form of soft X-ray stellar flares during the transient 
plasma processes in the stellar coronae (e.g., Pandey \& Singh, 2008).
Figure 1 shows the X-ray active stars detected along the H-R diagram 
(Huensch et al., 1997, 1998; Lalitha, 2013) based on ROSAT observations. The yellow circle 
represents the Sun that is also having X-ray active corona.

If we consider between the F-M spectral classes of the main-sequence stars, then there are significant detection 
of the cool and magnetically active stars which are having their X-ray active 
coronae associated with various transient activities (cf., Figure 1), e.g., flares and associated
plasma processes.
The magnetic field generation in the stars from F to mid-M spectral classes may be due 
to Parker type dynamo (1955) that generates the seed magnetic field further fanning out into their coronae to make them 
magnetically active. However, stars of spectral type M3 and others in the decreasing 
temperature down the main sequence
in H-R diagram
may be fully convective (Chabrier \& Baraffe 1997).
These cool dwarfs may not undergo the same dynamo process  
(Browning, 2008), therefore, they can be low X-ray luminous and active (Robrade \& Schmitt 2005)
stars during the typical transient activities. 
These stars can also produce strong impulsive flares of short duration as well as long duration gradual 
flares in their coronae very similar to the typical solar flares (Benz 2008; Fuhrmeister et al., 2008, 2011; Robrade et al., 2010). Therefore, there must be some
physical processes (e.g., typical reconnection process) in such stellar coronae of mid- to late-M type stars to build-up and release 
of the large amount of magnetic energy
stored in their coronal magnetic fields (e.g., Reiners \& Basri, 2010). 
Above mentiond these properties of the cool dwarfs are summarized in Fuhrmeister et al. (2011).
In conclusion, these Sun-like stars have seed magnetic field generation 
in their interior, which transport outward  to permeate their chromosphere and coronae 
in leading to the formation of transients (e.g., flares) as well as some form of MHD wave activity 
(e.g., Pandey \& Srivastava, 2009) similar to the transients and waves of the solar corona. As said above, these MHD wave 
modes are worth to diagnose the stellar coronae in the context of their localized physical properties.


\begin{figure}
\centerline{\includegraphics[width=10cm]{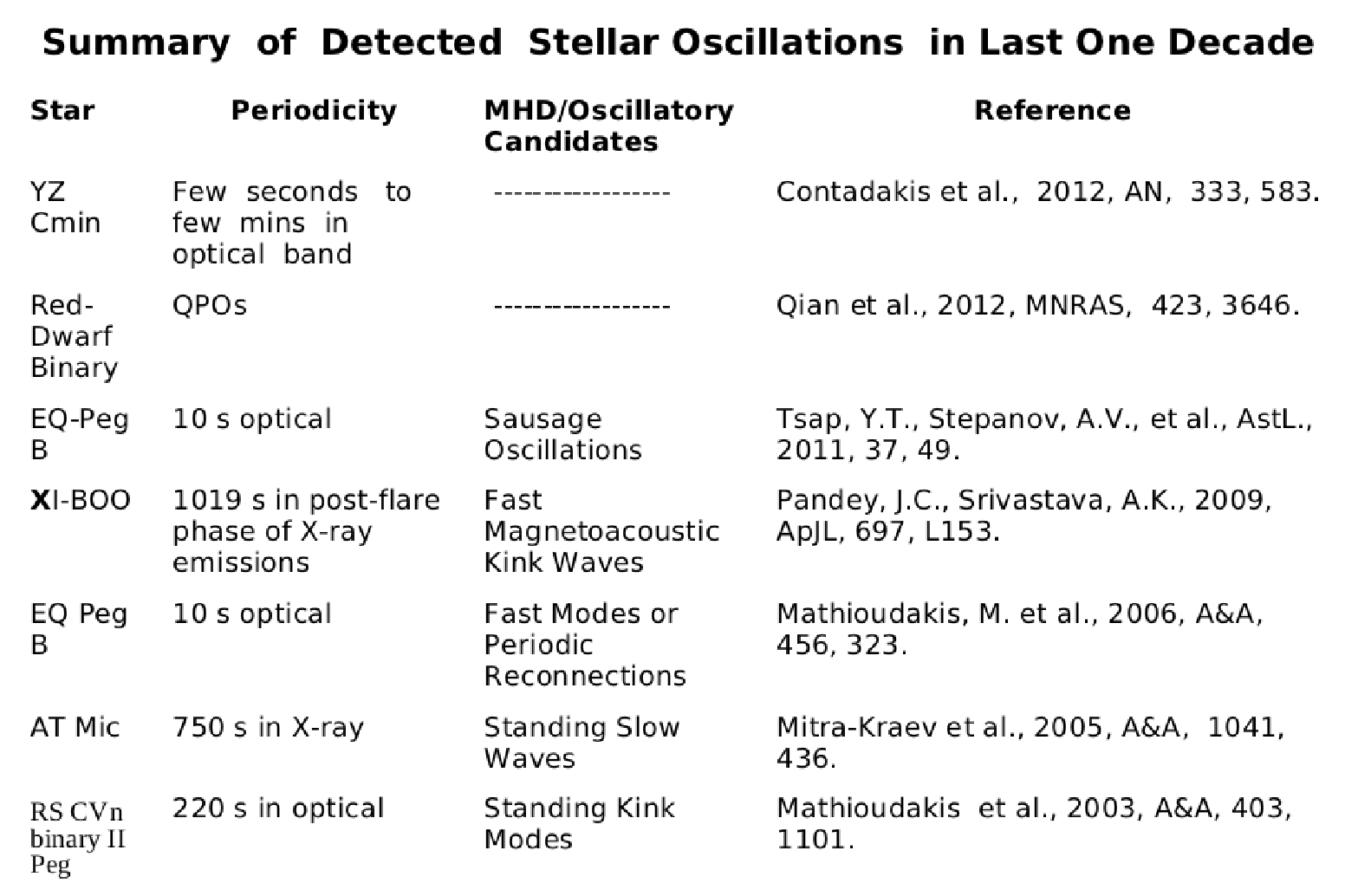}}
%
%
\end{figure}

\section{Detection of MHD Waves and Oscillations in Some X-Ray Active Sun-like Flaring Stars}\label{X-ray  Stars}

Few potential cases of the detection of stellar oscillations in magnetized stars are 
reported in optical as well as X-ray 
emissions as outlined in Table 1 since last one decade . 
However, there are only very few detection of the MHD waves and oscillations in the coronae of 
X-ray active Sun-like stars. Based on the solar analogy, the detection of such 
quasi-periodic oscillations can be a useful tool to determine the local plasma 
conditions of the stellar coronae (e.g., Nakariakov \& Verwichte,2005; Nakariakov \& Melnikov, 2009;
Srivastava,2010, and references cited there). Some of these potential findings in the context of stellar
oscillations are summarized 
as follows :

{\it Detection of slow magnetoacoustic waves in AT Mic by Mitra-Kraev et al. (2005)}

This is the first X-ray observation of an oscillation during a stellar flare in the corona of an active M-type dwarf AT Mic as observed by XMM-Newton
on 16 October 2000. The soft X-ray light curve (0.2-12 keV) is investigated that exhibits an evidence for a damped oscillation with a period of around 750 s with a damping time of 2000 s. It was claimed that the oscillation is a first evidence of standing magneto-acoustic waves generated by the impulsive loop heating during the flare. The mode was detected as a longitudinal, slow-mode standing wave, with a resulting loop length of (2.5 $\pm$ 0.2)$\times$10$^{10}$10 cm. The local magnetic field strength is estimated as 105 G
in the flaring loops. 

{\it Detection of fast magnetoacoustic kink waves in $\xi$-BOO by Pandey \& Srivastava (2009)}

The observations of X-ray oscillations during the flare in a cool active star $\xi$-Boo is reported for the first time as observed by EPIC/MOS of the XMM-Newton satellite
during the epoch on 19 January 2001. The X-ray light curve is investigated with wavelet and periodogram analyses, which show oscillations of the period of 1019 s  interpreted as a fundamental fast-kink mode of magnetoacoustic waves. These oscillations can be highly dispersive 
and can be a useful candidate for the localized heating of the corona of $\xi$-Boo. It should be noted that 
it is a binary K-Dwarf with the magnetized atmosphere and flaring activities.

{\it Observations of the decaying long period oscillation of a stellar megaflare by Anfinogentov et al. (2013)}

In the decay phase of the U-band light curve of a stellar megaflare observed on 2009 January 16 in the corona of dM4.5e star YZ CMi,
the long-period oscillations of 32 min with a decay time of 46 minutes is observed with its most likely
interpretation as the impulsive generation of slow magnetoacoustic oscillations in the stellar loop system.

{\it Detection of multiple slow magnetoacoustic oscillations in Proxima Centauri by Srivastava et al. (2013)}

The first observational evidence of multiple slow acoustic oscillations in flaring loops in the corona
of Proxima Centauri using the XMM/Newton observations of 14 March 2009 has been detected. The oscillations in the decay phase of the flare 
in its soft X-ray emissions (cf., Figure 3) are found with the detection of multiple periodicities
of 1261 s and 687 s respectively with the probability of $>$99 \% and for $>$ 4 cycles.
. The appearance of such oscillations are bursty during flare energy release, and exhibit
strong decaying nature 
similar to the slow acoustic oscillations observed in the solar corona (e.g., Wang et al., 2002).
Thermal conduction
may also play a significant role in the dissipation of observed slow waves in the stellar corona of
Proxima Centauri. The period ratio P$_{1}$/P$_{2}$ is found to be below than 2.0, i.e., 1.83,
which may be the signature of the longitudinal density stratification of the stellar loops in which such oscillations
are excited. Using the half loop length of 7.5 $\times$ 10$^{9}$ cm, and period ratio, the 
density scale height of the stellar loop system is estimated as 23 Mm based on the analogy of solar corona (e.g., McEwan et al.,2006;
Srivastava et al., 2010; Macnmara \& Roberts, 2010; Kumar et al., 2011; Luna-Cardozo et al., 2012, and references cited therein). The density scale height is well below the hydrostatic scale
height of such loops assuming the few mega-Kelvin temperature, which indicates the existence of non-equilibrium conditions there, e.g., flows and mass
structuring. 
First clues of the excitation of  
multiple slow acoustic oscillations in the hot stellar loops in the corona of Proxima Centauri
are obtained and its MHD seismology is performed to diagnose the local
plasma conditions (Srivastava et al., 2013). We outline the summary 
of this new important detection published in the Astrophysical Journal Letters in 2013.


\begin{figure}
\centerline{\includegraphics[width=11cm,angle=90]{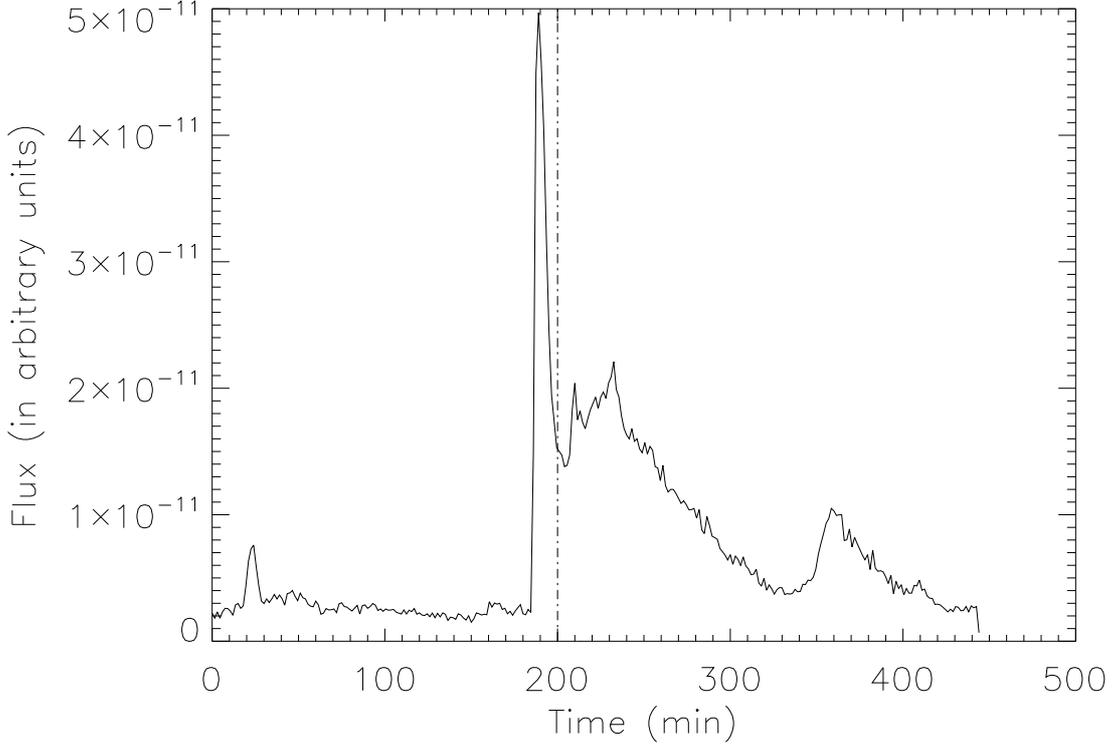}}
\caption{XMM-Newton light curve of a flaring epoch in the corona of Proxima Centauri on 14
March 2009. The vertical dotted line shows the time when quasi-periodic oscillations trigger-on in
the decay phase of the flare. The oscillatory light curve in the decay phase of the flare has been
chosen for the detection of the significant periodicities in the soft X-ray emissions.
The periodicities have been detected using the wavelet analyses on the 
de-trended light curve after the removal of long-term variation.}
\label{f:one}
\end{figure}

\section*{Discussions and Conclusions}

The heightened and integrated X-ray emissions from the un-resolved 
coronae of the distant Sun-like stars during the flaring activity 
give some episodic quasi-periodic pulsations especially in the 
post-flare phases (e.g., Pandey \& Srivastava, 2009). The reconnection 
in the formed current-sheets in the stellar coronae may dissipate 
the magnetic energy and generate the heating, energetic particles, as well as  shocks
(e.g., Nakariakov \& Melnikov, 2009; Shibata \& Magara, 2011). The flare generated 
disturbances may also trigger the fast magnetoacoustic waves (e.g., kink, sausage)
in the nearby stellar loops similar to the oscillations of coronal loops 
(e.g., Srivastava et al., 2008; Aschwanden \& Schrijver, 2011, and references cited therein).
The longitudinal sausage oscillations can directly modify the density and thus 
the intensity, which may generate the X-ray QPOs emitted from the stellar coronae
(e.g., Antolin \& Van Doorsselaere, 2013). While, the kink oscillations in the tangled 
stellar loops w.r.t. L.O.S. can also cause the variation in the plasma column depth, 
and thus can generate the weak density perturbations as well as intensity 
oscillations in the stellar flare emissions (Cooper et al., 2003). The strong 
energy release can also trigger the non-linear kink oscillations that can perturb
the density and can generate the weak intensity oscillations (Andries et al., 2009).
Therefore, the quasi-periodic oscillations can be triggered as a proxy of these fast wave modes
in various emissions from the stellar loops during the flare epoch.
Apart from the fast mode waves, the impulsive foot-point heating in the stellar loops during the flare energy 
release may generate the slow-acoustic oscillations similar to the slow 
modes of the solar coronal loops (e.g., Wang, 2010, and references cited therein).

In conclusion, the strong flares that are being frequently observed in the Sun-like flaring stars (Maehara et al., 2012), may be enough energetic to trigger the various magnetohydrodynamic (MHD) wave modes.
The detection of such wave modes in the stellar coronae may give potential 
information on the physical processes that generate such waves, as well as can 
be utilized to understand the local plasma conditions there (e.g., Mitra-Kraev et al., 2005;
Pandey \& Srivastava, 2009; Srivastava \& Lalitha, 2013, and references cited therein). The future space-borne observational
campaign should be planned for observing more samples of the flaring epoch of Sun-like stars 
to achieve significant statistics about the detection of the MHD oscillations in their coronae, and therefore, the 
diagnostics of local plasma properties in the framework of MHD seismology can be performed more convincingly keeping the view of solar analogy.

\section*{Acknowledgements}
We acknowledge the valuable comments of the referee
that improved our manuscript.
AKS acknowledges the support of DST-RFBR (Indo-Russian) Project 
(Ref : INT/RFBR/P-117) during the research  work.
AKS thanks Shobhna Srivastava for the patient encouragements.


\end{document}